\documentclass[]{spie}  

\usepackage{amssymb,amsmath,mathabx} 
\usepackage{multirow}
\usepackage{graphicx, float}
\usepackage[euler]{textgreek}
\usepackage[colorlinks=true, allcolors=blue]{hyperref}

\title{A precision cryogenic positioning stage for detector dithering and flexure compensation}

\author[a]{Stephen A. Smee}
\author[a]{Stephen C. Hope}
\author[a]{Randolph P. Hammond}
\author[a]{Albert C. Harding}
\author[b]{Tyson Hare}
\author[a]{Aidan C. Gray}
\author[a]{Katherine G. Smee}
\author[b]{Leon Aslan}
\author[c]{Robert H. Barkhouser}
\author[d]{Andrea Bianco}
\author[b]{Christoph Birk}
\author[b]{Maren Cosens}
\author[d]{Michele Frangiamore}
\author[b]{Daniel D. Kelson}
\author[b]{Gerrad Killion}
\author[b]{Nicholas P. Konidaris II}
\author[e]{Alicia Lanz}
\author[a]{Jacob McCloskey}
\author[b]{Andrew B. Newman}
\author[b]{Solange Ramirez}
\author[b]{Gwen C. Rudie}
\author[d]{Andrea Vanella}
\author[b]{Jason E. Williams}

\affil[a]{Johns Hopkins University, Department of Physics and Astronomy, 3701 San Martin Drive, Baltimore, MD 21218, USA}
\affil[b]{Carnegie Observatories, 813 Santa Barbara Street, Pasadena, CA 91101, USA}
\affil[c]{LCS Optics LLC, Parkton, MD, USA}
\affil[d]{INAF -- Observatorio Astronomico di Brera, via Bianchi 46, 23807, Merate, Italy}
\affil[e]{Capella Space, 438 Shotwell St., San Francisco, CA, 94110}

\authorinfo{Further author information: (Send correspondence to Stephen Smee)\\E-mail: smee@jhu.edu, Telephone: 1 410 516 7097}

\begin{document}  
\maketitle

\begin{abstract}
This paper presents the design and technical progress of a precision X-Y stage for detector dithering and flexure compensation.  The stage is being developed for use in the Magellan InfraRed Multi-Object Spectrograph, MIRMOS.  MIRMOS is a very large Nasmyth mounted spectrograph containing a combination of refractive, reflective and diffractive optics mounted on a long cryogenic optical bench.  The instrument utilizes five science cameras, each having a custom x-y stage to control the in-plane detector position within each camera, providing both dithering capability for improved sampling, and flexure compensation to correct for image motion that results from the gravity variant operation of the instrument.  Designed to operate at 120~K, the stage will accurately control detector position in two orthogonal degrees of freedom, and have manual fine adjustment features to set detector tip, tilt and piston.  The piezo-driven flexure stage provides high-resolution backlash-free motion of the detector and is very compact along the optical path, keeping camera length to a minimum.   A magnetoresistive bridge provides position feedback in each degree of freedom, greatly reducing hysteresis, which is common in piezoelectric actuators.  The system is designed to operate in open loop using a lookup table keyed to the Nasmyth rotator angle for flexure control.  Here, the optomechanical design of the stage, electrical control system, and current performance results from early prototype efforts are presented and discussed. 
\end{abstract}


\section{INTRODUCTION}
\label{sec:intro}

Currently under development, the Magellan Infrared Multi-Object Spectrograph (MIRMOS) will be a workhorse instrument for the Nasmyth port of the 6.5-meter Magellan telescope at Las Campanas Observatory~\cite{2022SPIE12184E..15K}.  MIRMOS collects spectra in the infrared, operating in the Y (0.886~$\mu$m – 1.124~$\mu$m), J (1.124~$\mu$m – 1.352~$\mu$m), H (1.466~$\mu$m – 1.807~$\mu$m), and K (1.921~$\mu$m – 2.404~$\mu$m) bands.  Full bandpass spectra are collected in a single shot, utilizing four H2RG detectors, one per band.  Volume Phase Holographic (VPH) gratings disperse light in each band, at medium resolution, with resolving power R = $\lambda/\Delta\lambda$ = [3,400, 4,000, 3,800, 3,600] in Y, J, H, and K, respectively.  A cryogenic slit mechanism will allow rapid reconfiguration of slits within the focal plane.  In multi-slit mode, MIRMOS will cover the $13'\times 3'$ field of view with 92 slits.  In integral-field mode, MIRMOS will cover a $26"\times 20"$ field with 1" wide slices. In diffuser-assisted mode, a diffuser will be inserted into the beam to allow for $R\sim300$ differential spectroscopy with unprecedented spectrophotometric precision.

\begin{figure}
    \centering
    \includegraphics[width=0.9\textwidth]{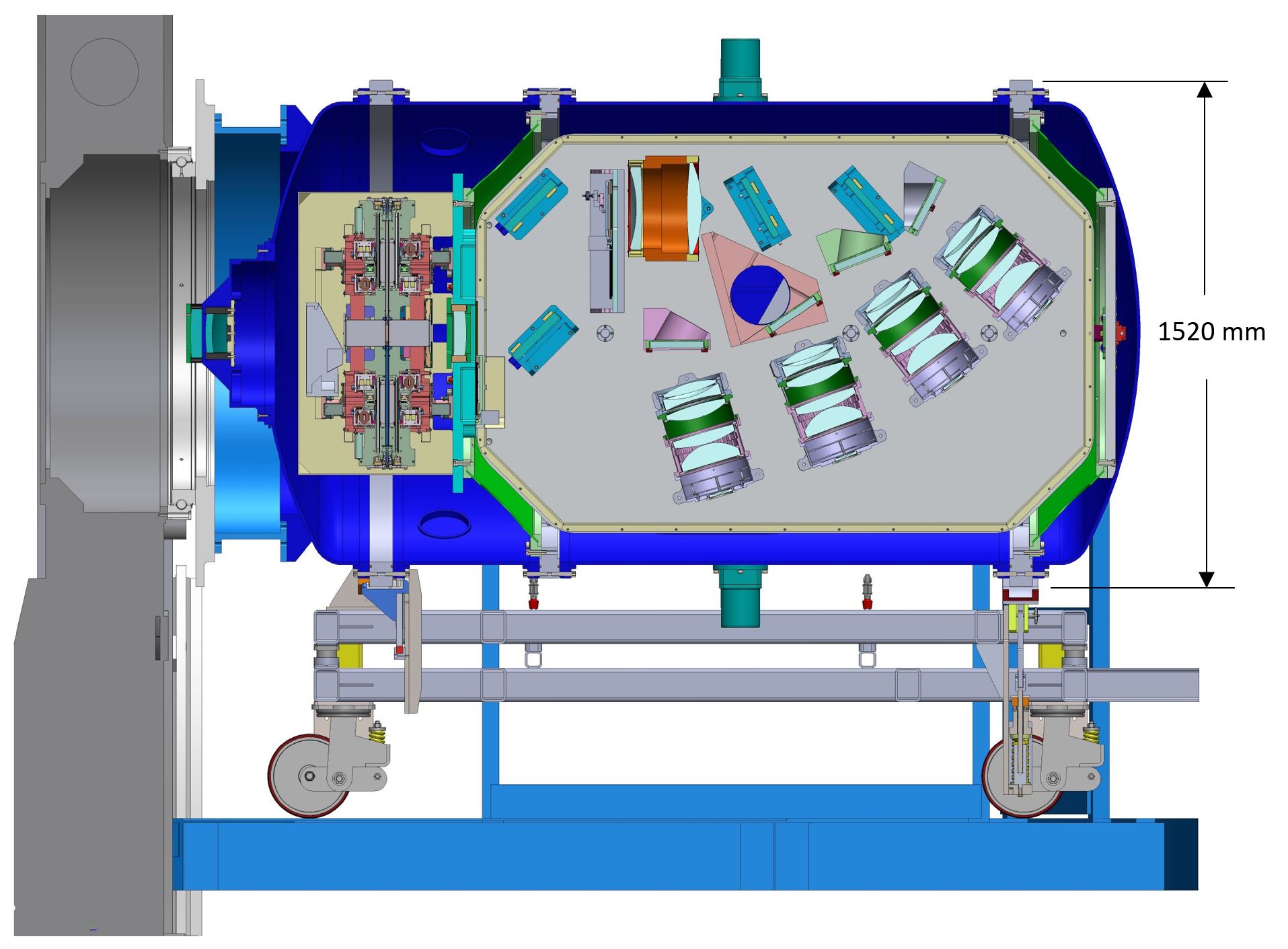}
    \caption{MIRMOS mounted on the Magellan Nasmyth platform.  Light enters from the left, through the vessel window, and propogates through the optical train, ultimately producing spectra on the science detectors,.  The instrument rotates about the vessel axis to counter field rotation as the telescope slews. A cart with astatic rollers support the rear of the vessel, reducing load on the rotator bearing.}
    \label{fig:MIRMOS_Nasmyth}
\end{figure}

MIRMOS is a large vacuum-cryogenic instrument operating in a gravity-variant environment and will be subject to flexure induced image motion at the science detectors.  The cryostat vessel is roughly 1.5~m in diameter and 3~m long; see Figure~\ref{fig:MIRMOS_Nasmyth}.  The instrument optical axis is coincident with the both the mechanical axis of the vessel and the elevation axis of the telescope.  Between MIRMOS and the telescope is an instrument rotator that compensates for field rotation as the telescope tracks the sky.  During observations, rotation of the instrument about the elevation axis will induce flexure due to the changing gravity vector, producing a time varying motion of the spectra on the science detectors.  This reduces spectral resolution since motion of the spectra during observations, typically $\sim$300 seconds long, smears the science signal over potentially many pixels.  Additionally, for accurate sky subtraction, sky lines should remain stationary to roughly 1/15 of a pixel, of order 1~$\mu$m.  Therefore detector position must be stable, more stable than can reasonably be expected from a passive design for an instrument of this scale.  Automated detector position control is needed to enable accurate sky subtraction, and that control will require a resolution of some reasonable fraction of a micron to mitigate degradation in performance due to flexure.   In addition, the range of motion should be suitable to correct flexure over the maximum envisioned slew of the rotator during an exposure, considering the stiffness of the instrument structure and optomechanics. Based on rudimentary analysis, the anticipated range of motion needed is not expected to exceed $\sim200~\mu$m in the X and Y axes.

A separate, but coupled, consideration exists with regard to spectral sampling.  MIRMOS has an image quality requirement of 0.35" FWHM, sufficient to take advantage of the superb seeing conditions that occur at Magellan.  However, the equivalent-light-gathering-power of the science cameras is f/1.4 leading to a plate scale of $\sim$0.4" per pixel (when averaging spatial and spectral directions), well below the Nyquist limit for sampling. To correct for this limitation, half-pixel dithering will be used to recover this otherwise lost image quality. 

This paper presents the design of a piezo-driven X-Y detector stage for flexure compensation and dithering.  The Dithering and Flexure Compensation System (DiFCS) developed for MIRMOS will operate in a vacuum and at cryogenic temperature.  In the sections that follow we present the DiFCS design and discuss the results collected to date from a prototype device.  

\section{DESIGN}
\label{sec:design}

MIRMOS is a large cryogenic instrument with a cold operating temperature of 120~K, and a detector working temperature of 80~K.  It is long, just fitting within the size limitations posed by the width of the instrument lift platform and depth of the Nasmyth platform.  To minimize the size of the vacuum vessel the science cameras have been designed to be as short as reasonably possible without impacting performance.  The mechanical stage system used to correct for gravity-induced flexure, and to provide dithering capability, must be compact along the camera axis to minimize the diameter and length of the instrument.  Furthermore, it must provide micron-level resolution in both the spatial and spectral directions of the detector with a range of motion of $\sim$ 200~$mu$m. The DiFCS design satisfies these basic physical and operational requirements and is discussed here in more detail.

\subsection{The MIRMOS Cameras}
\label{sec:camera}

\begin{figure}
    \centering
    \includegraphics[width=0.99
    \textwidth]{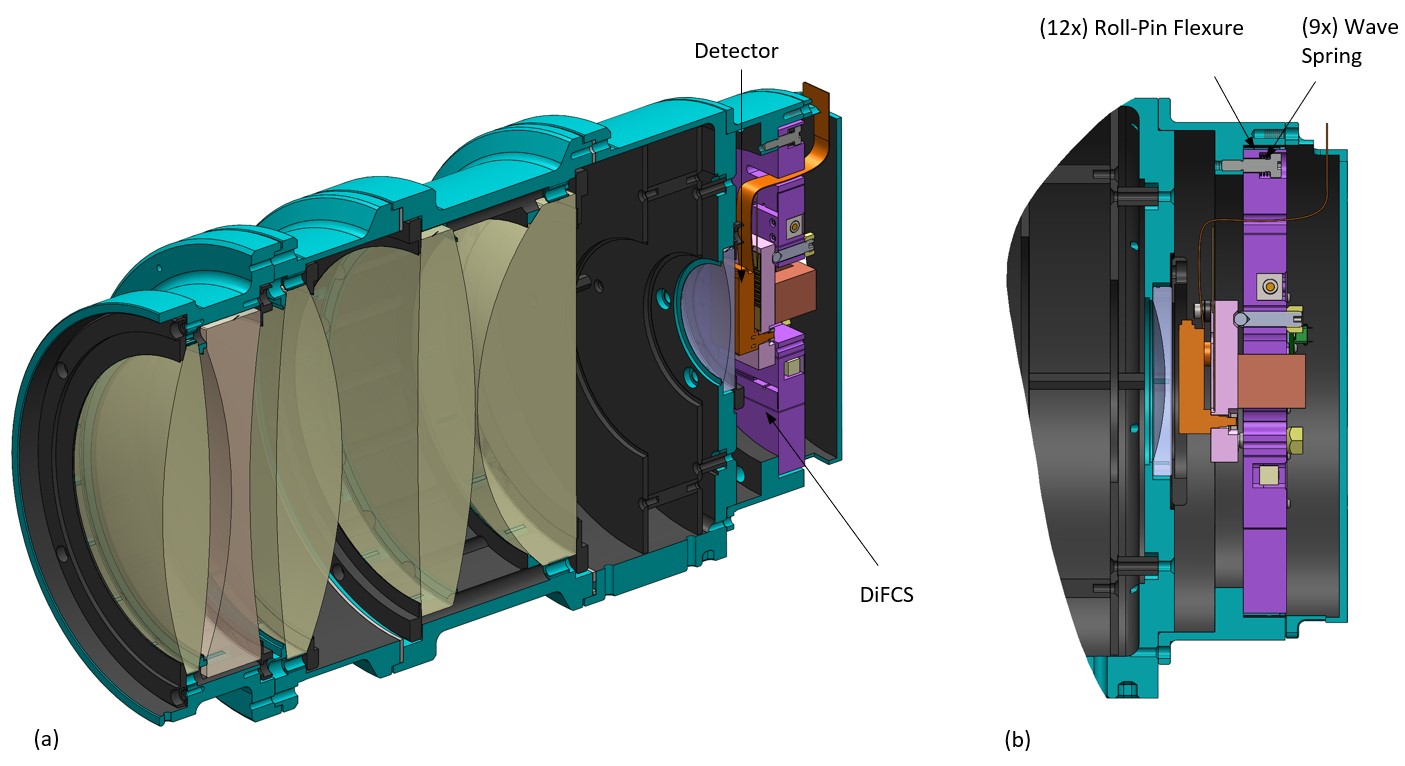}
    \caption{Section view of the MIRMOS J-Band Camera.  The six-element all refractive design focuses dispersed light onto an H2RG mounted on the DiFCS Stage. (a) Isometric view.  (b) Detailed section view highlighting the DiFCS attachment to the rear of the camera.}
    \label{fig:MIRMOS_camera}
\end{figure}

MIRMOS utilizes one visible bandpass camera for slit viewing and science imaging, and four near infrared cameras of almost identical design; see Figure~\ref{fig:MIRMOS_camera}.  All of these cameras are refractive six element designs that focus incident light onto a science detector; a CCD in the case of the visible camera, and an H2RG in the case of the four near infrared cameras.  Lens elements are mounted in aluminum cells supported by roll-pin flexures for very precise alignment~\cite{2010SPIE.7739E..3OS}.  Like the lenses, the DiFCS is also mounted in a roll-pin flexure cell, residing just behind the image plane.  The DiFCS supports the detector, facilitates manual adjustment for tip-tilt-piston during integration, and controls the in-plane position of the detector during operation for flexure compensation and dithering.

\subsection{The DiFCS X-Y Stage}
\label{sec:X-Y}
\begin{figure}
    \centering
    \includegraphics[width=1.0\textwidth]{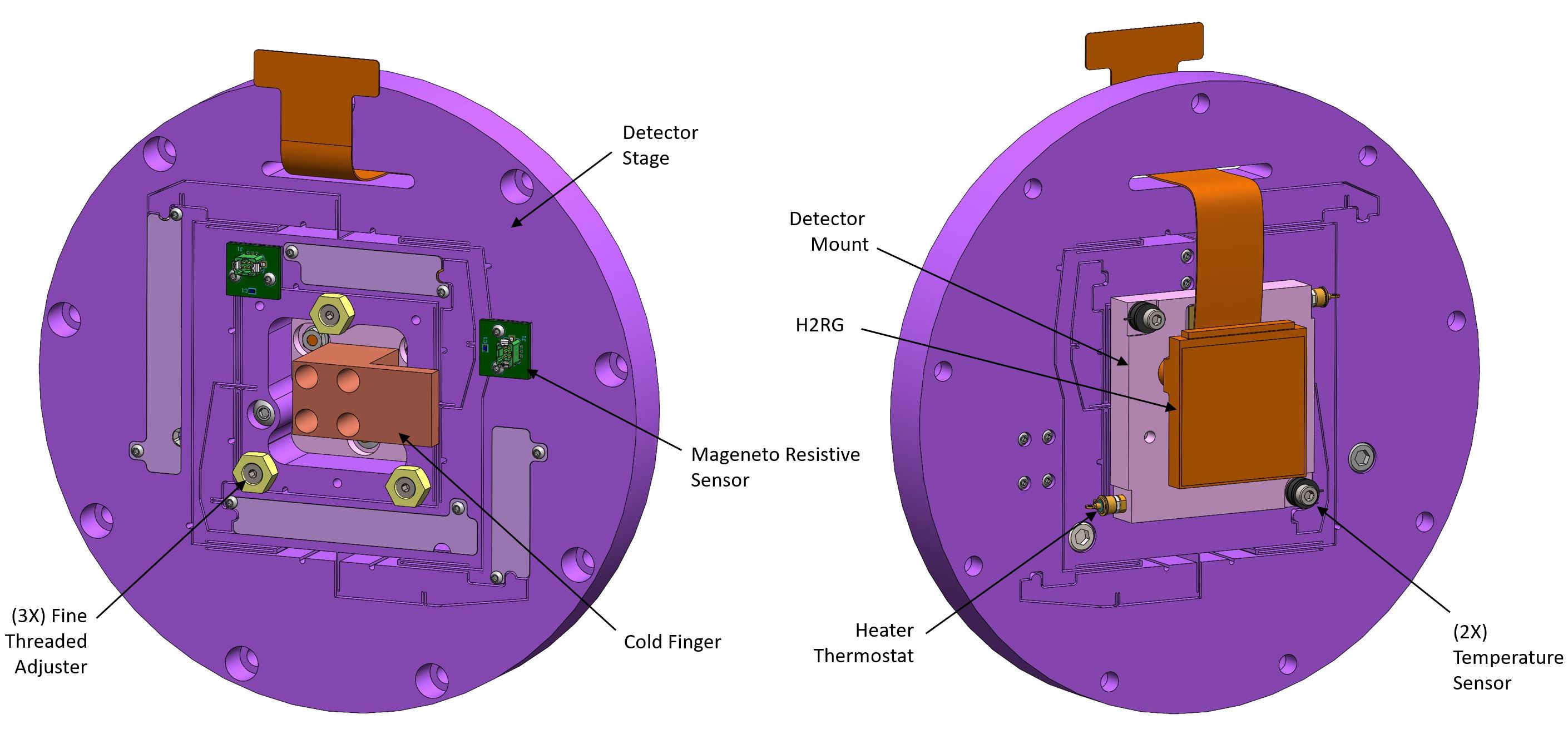}
    \caption{The Dithering and Flexure Compensation System.  a) Rear view showing the rear of the DiFCS. b) Front view, showing the detector mounting stage and the detector.}
    \label{fig:DiFCS_ISO}
\end{figure}

\begin{figure}
    \centering
    \includegraphics[width=0.85\textwidth]{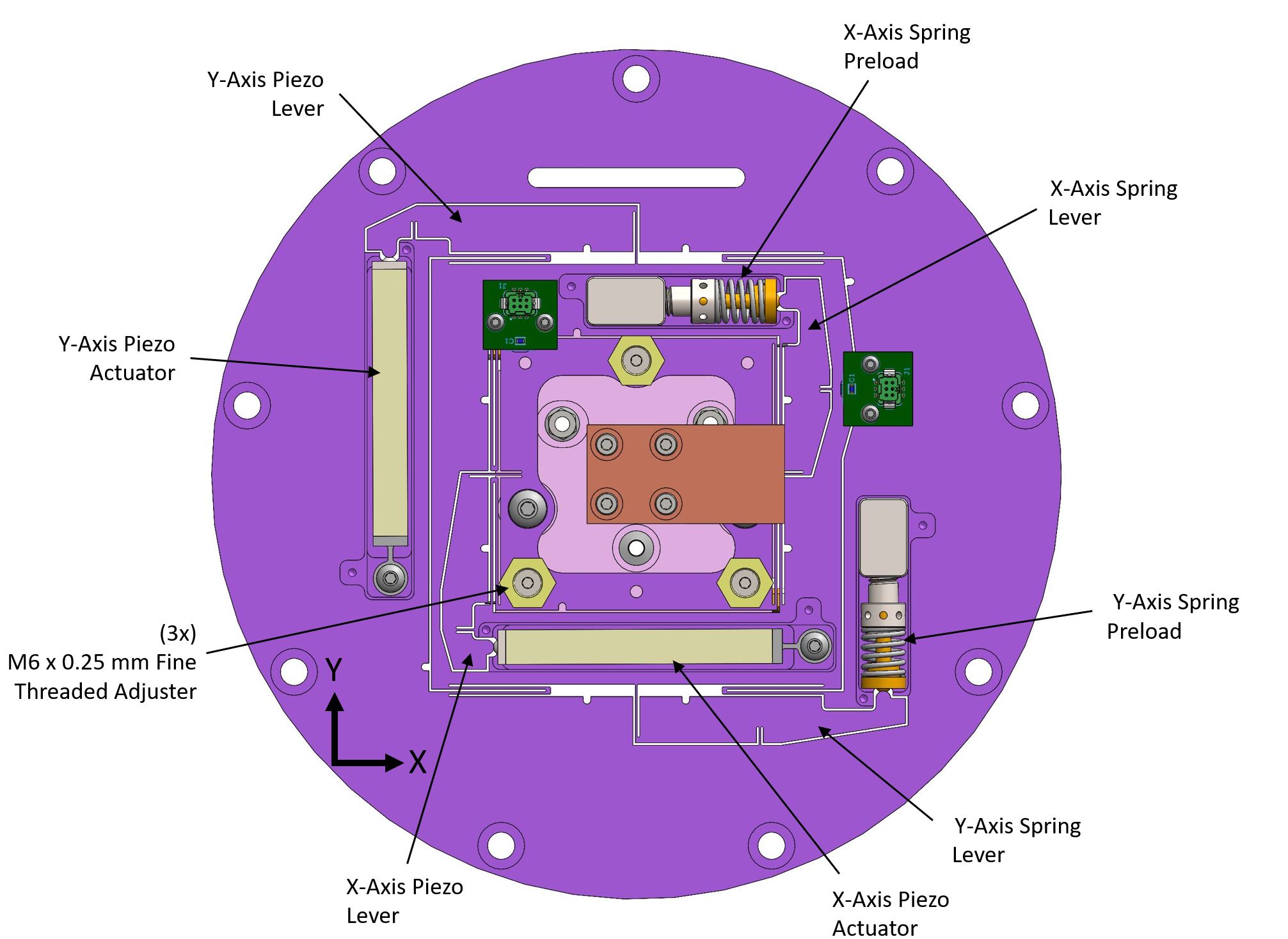}
    \caption{Rear view of the Dithering and Flexure Compensation Stage (DiFCS) showing the piezoelectric actuators for the X-Y degrees of freedom, the actuator preload springs, and the three fine-threaded adjusters for tip/tilt/piston adjustment of the detector.}
    \label{fig:DiFCS_Rear}
\end{figure}

The piezo-driven DiFCS X-Y flexure stage, shown in Figure~\ref{fig:DiFCS_ISO}, is a relatively thin (16~mm) titanium (6Al4V) plate that supports the detector on the front side and integrates the piezos drive and positional encoding into the rear of the plate.  Flexures integrated into the plate provide backlash-free motion in the spectral and spatial (for convenience, X and Y) directions; see Figure~\ref{fig:DiFCS_Rear}.  Each degree of freedom is driven by a piezo-electric actuator through a 7:1 lever to compensate for the reduced stroke of the piezo actuator when operating at $\sim$120~K. An adjustable spring preload ensures solid contact between the piezo tip and the lever over the entire range of travel. Positional encoding using a magnetoresistive sensor is also provided in each degree of freedom.  In operation, a voltage is applied to each piezo element in proportion to the stroke required.  The system will operate in an open-loop mode with positions read from a lookup table generated by characterizing flexure for each camera as a function of the Nasmyth rotator angle.  The DiFCS stage itself will operate in a closed-loop mode using the magnetoresistive sensor to determine position location.  In short, stage axis position maps to Nasmyth rotator angle, and the correct voltage is applied to the piezo element to achieve that position; using the magnetoresistive sensor alleviates concerns about hysteresis in the piezo.

\subsection{The Piezo Actuator}
\label{sec:piezo}

The piezo actuator consists of a piezoelectric element with a metal (Alloy 42) tip bonded to each end. These tips provide the mechanical interface to the titanium flexure stage.  The piezoelectric element is provided by American Piezo and consists of three bonded piezo ceramic elements, model 45-1150 PSt150/7x7/20, having overall dimensions of 7~mm $\times$ 7~mm $\times$ 54~mm.  Total stroke for the piezo stack is 60~$\mu$m at room temperature.  With the 7:1 mechanical advantage integral to the stage design the total travel in each axis at room temperature is 420~$\mu$m.  At the operating temperature of 120~K we expect the piezo stack to produce only 50\% of the room temperature stroke, which equates to a stage motion of $\sim$210~$\mu$m, providing slightly more than the target range of 200~$\mu$m.

\begin{figure}
    \centering
    \includegraphics[width=0.85\textwidth]{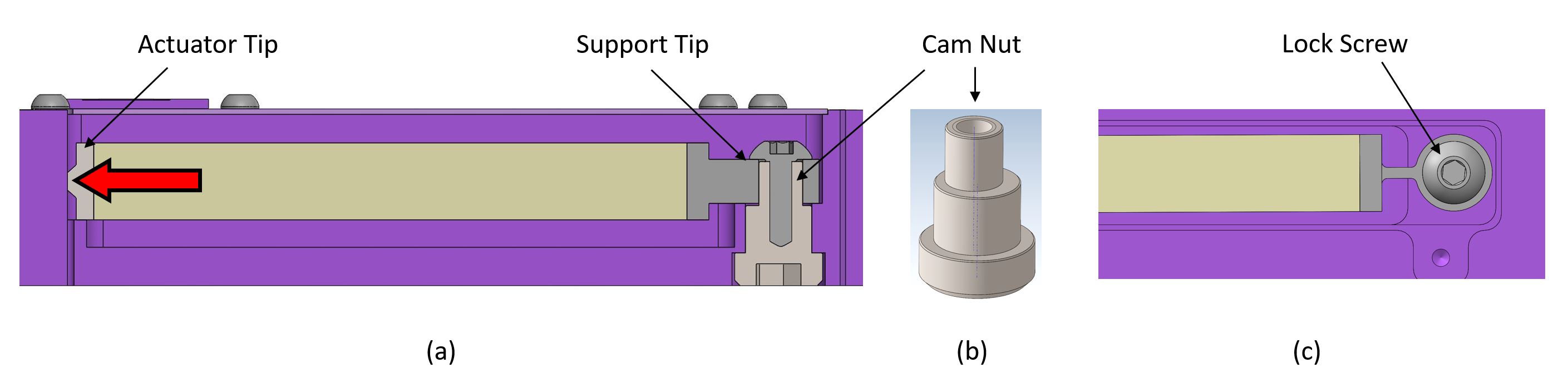}
    \caption{(a) Side view cross-section showing the piezo actuator and interface to the stage.  (b) Rendering of the cam nut.  (c) Top view highlighting the cam nut feature.}
    \label{fig:DiFCS_Cam}
\end{figure}

It is important to note that piezo elements are ceramic and have a low tolerance for tensile loading.  This has some ramification for the design as it relates to the stage flexures. To minimize stress in the flexures it is desirable to set the range of motion to operate about the neutral position of the stage, where zero applied voltage corresponds to the maximum negative travel (-100~$\mu$m when cold) and the maximum applied voltage corresponds to the maximum positive travel (+100~$\mu$m when cold).  This necessitates the spring preload.  Too tune the zero position to the zero voltage position a cam nut is used to adjust the axial position of the fixed-end of the piezo actuator; see Figure~\ref{fig:DiFCS_Cam}.  The feature also compensates for manufacturing tolerances in the stage and actuator.

\subsection{Detector Tip/Tilt/Piston}
\label{sec:ttp}
\begin{figure}
    \centering
    \includegraphics[width=0.8\textwidth]{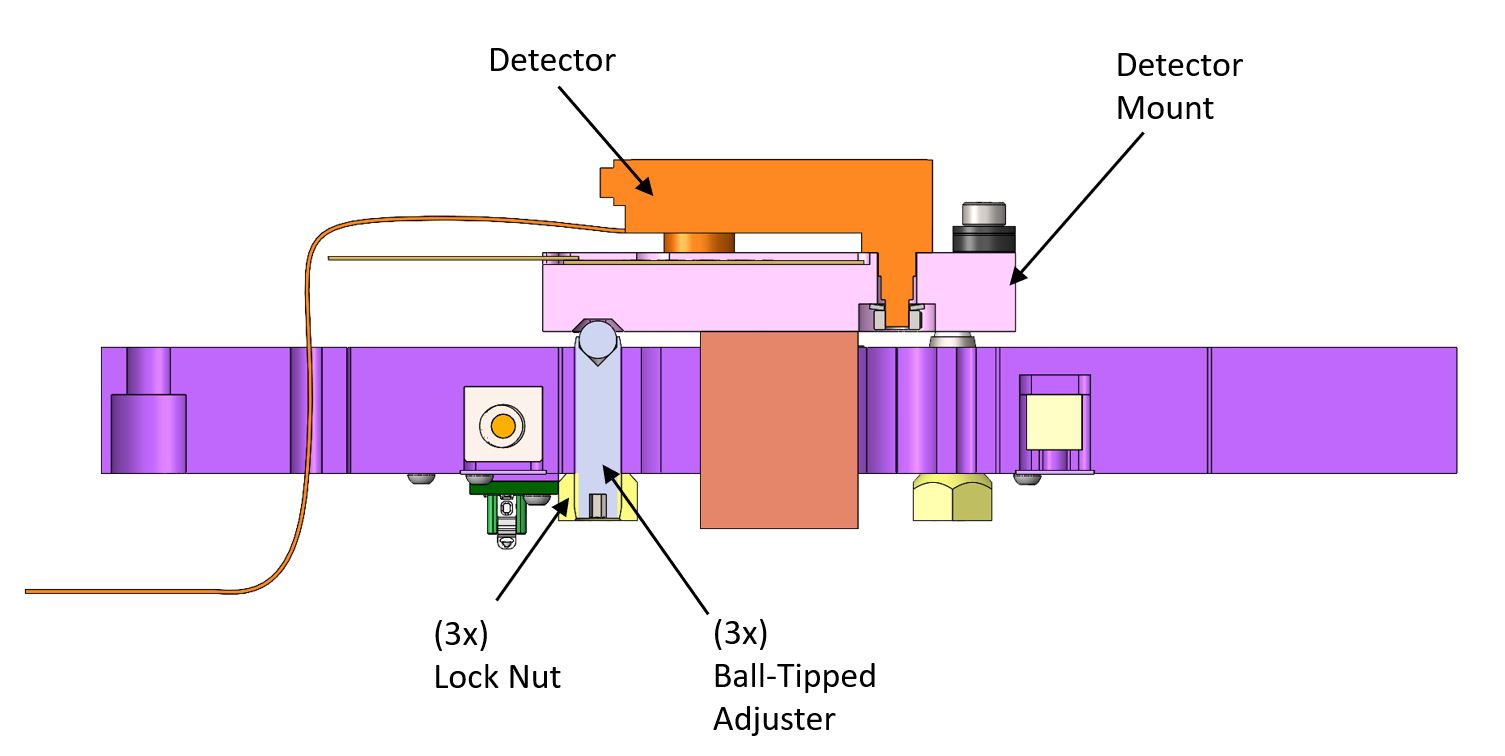}
    \caption{Section view of the DiFCS highlighting elements of the tip/tilt/piston adjustment.  Three fine thread ball-tipped adjusters interface to corresponding v-grooves machined into the rear of the detector mount plate.  The adjustments can be performed from the rear of the DiFCS without need to remove the detector prior to adjustment.  A hex not locks each adjustment.}
    \label{fig:tiptilt}
\end{figure}

Optomechanical component tolerances within MIRMOS require a set-and-forget detector focus adjustment to achieve best-focus for the as-built cameras. Additionally, the MIRMOS focal plane is slightly tilted, an optimization often used in spectrograph designs to optimize image quality across the detector surface.  To accommodate both considerations, the DiFCS implements a simple tip/tilt/piston adjustment scheme, identical to what was used for the WHIRC near infrared camera \cite{2011PASP..123...87S}; see Figure~\ref{fig:tiptilt}.  The detector mounting plate, made from molybdenum to match the thermal expansion coefficient of the H2RG detector, is supported kinematically by three M6 $\times$ 0.25~mm fine-threaded ball-tipped adjusters. Each adjuster, accessible from the rear of the DiFCS, integrates a hardened tooling ball at the adjuster tip which engages a V-groove machined into the rear of the molybdenum detector mount plate.  A lock nut on each adjuster secures the adjustment.  Two compression springs buried in the rear of the DiFCS preload the tooling balls against the detector mount plate ensuring constant contact between the tooling balls and the V-groove surfaces.

\subsection{Thermal Considerations}
\label{sec:temp}
The simple tip/tilt/piston approach is elegant mechanically, but also elegant for thermal reasons.  The H2RG detectors in MIRMOS operate at a temperature that is colder than the MIRMOS optics, 80~K.  Insulating the detector from the surrounding 120~K environment happens naturally by virtue of the point contact between the tooling balls and the V-grooves in the molybdenum detector mount.  Additional thermal impedance is provided by the low thermal conductivity of titanium and the long thermal path through the bi-fold flexures.  Combined, the high thermal impedance allows the detector to be controlled by a separate cooling circuit requiring minimal lift and low parasitic conduction to the 120~K DiFCS.  In this design, as has been done for past instruments, a film heater integral to the detector mount plate is used to stabilize the detector temperature to high precision. Redundant thermal sensors are used for thermal sensing/control, and redundant thermostats guard against inadvertent overheating of the detector.

\subsection{Stage Drive Electronics and Motion Control}
\label{sec:electronics}

The DiFCs controller is designed to provide absolute detector positioning in X-Y space, allowing it to compensate for error induced by flexure in the optical bench, and to add detector dithering capability. The controller is fully integrated, consisting of: one isolated supply generating 160~V for the X-Y stage piezo drivers, and one isolated supply generating low voltages for the control electronics; an isolated RS485 interface for telemetry, command and control; a pair of high-voltage linear amplifiers, controlled by two 14 bit DAC’s, to drive the X-Y stage piezo elements; four 16 bit ADC’s providing X-Y positional feedback from a pair of sine-cosine magnetoresistive bridges; an accelerometer for angular rotation; and an embedded microcontroller.

Custom firmware provides control in several ways. In open-loop mode, the stage can be commanded with absolute voltages. This is useful for sensor calibration, where sensor output is calibrated to externally measured motion. Once calibrated, it can operate in a closed loop fashion, providing absolute positioning in microns. To provide flexure compensation a second calibration maps flexure induced X-Y image offset in microns to angular position of the Nasmyth port.  Finally, pre-programmed offsets can be added to provide precise detector dithering.

\section{PERFORMANCE}
\label{sec:perf}
 
Performance of the DiFCS has been investigated analytically and experimentally.  At the analytical level finite element analysis was used to predict material stress levels, stage stiffness, out-of-plane piston and tilt as a function of in-plane motion.  At the experimental level, stage motion has been characterized using precision metrology. Additional details are provided below. 

\subsection{Analytical Performance}
\label{sec:fea}

Flexure stress and out-of-plane motion were analyzed using finite element analysis, with the primary goal of understanding the stiffness and stress margin in the various flexures.  These flexures include: the "bi-fold flexures", which guide the linear motion in each degree of freedom; the "lever flexures", which act as a fulcrums for the piezo lever and spring preload lever; and the "pusher flexures", which connect the levers to the movable stage.  In all cases, stress levels were determined to be modest.  The maximum Von Mises stress occurs in the pusher flexure for the X axis, with a maximum of 170~MPa; low compared to the 880~MPa yield strength of titanium. 

\begin{figure}
    \centering
    \includegraphics[width=0.6\textwidth]{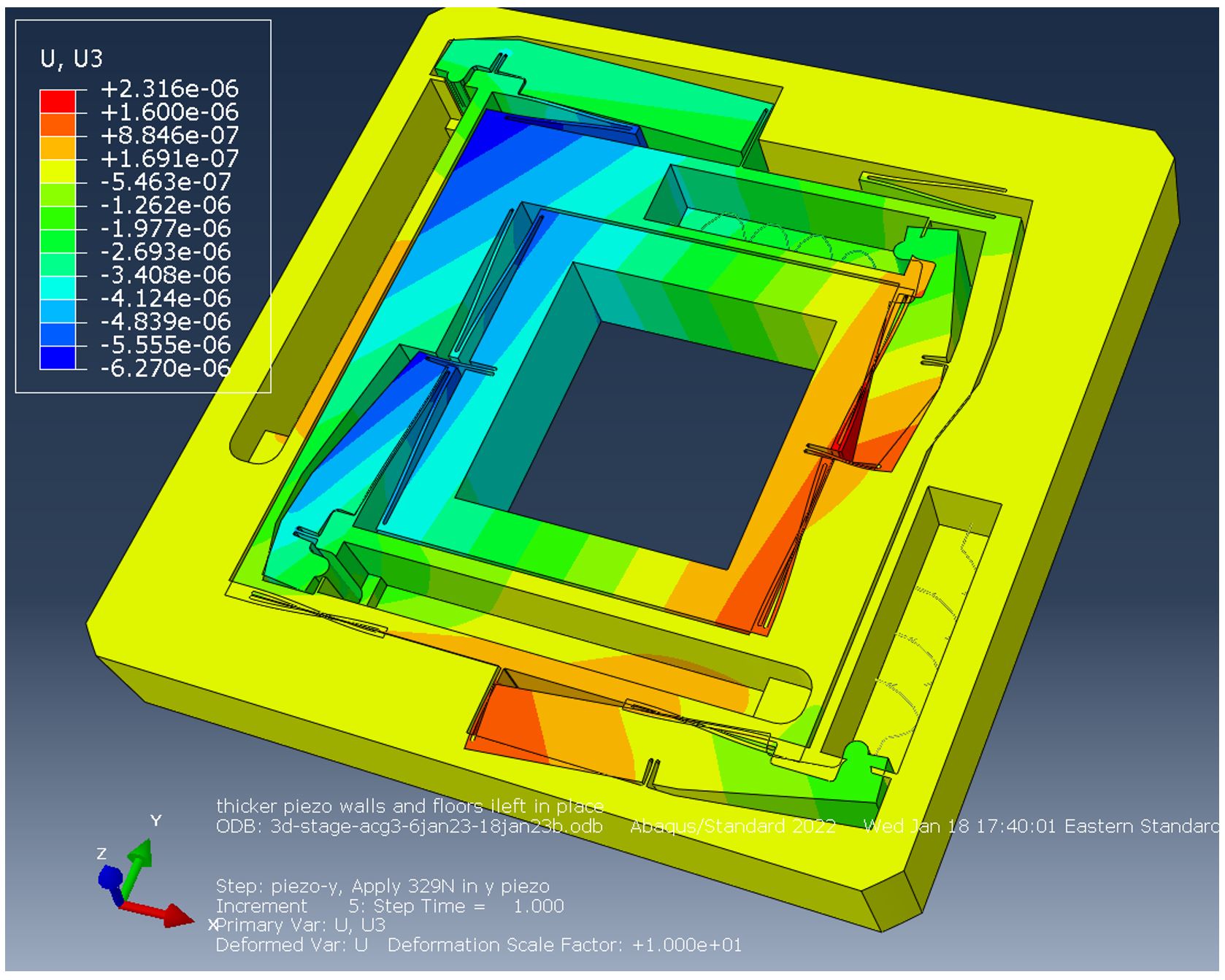}
    \caption{Out of plane deflection of the DiFCS stage due to in plane motion.  Results are shown for the maximum travel in X and Y.  U3 denotes the "Z" direction in this contour plot.  Note, this figure is presented with a deformation scale of 10X.}
    \label{fig:DiFCS_FEA}
\end{figure}

Out-of-plane motion (focus direction) was also analyzed, as this motion leads to defocus.  Analysis shows that the asymmetry in the design about the mid-plane (caused by the blind pockets that house the piezos and the preload springs) causes a slight out-of-plane motion (defocus and tilt) as the stage is slewed to the extremes of travel.  The tilt induced is small, $\sim$5~$\mu$m across the diagonal of the detector; see Figure~\ref{fig:DiFCS_FEA}.   Out-of-plane motion due to gravity is uniform and negligible.

\subsection{Measured Performance}
\label{sec:Perf}

\subsubsection{Ambient Performance}
\label{sec:amb_meas}

\begin{figure}
    \centering
    \includegraphics[width=0.99\textwidth]{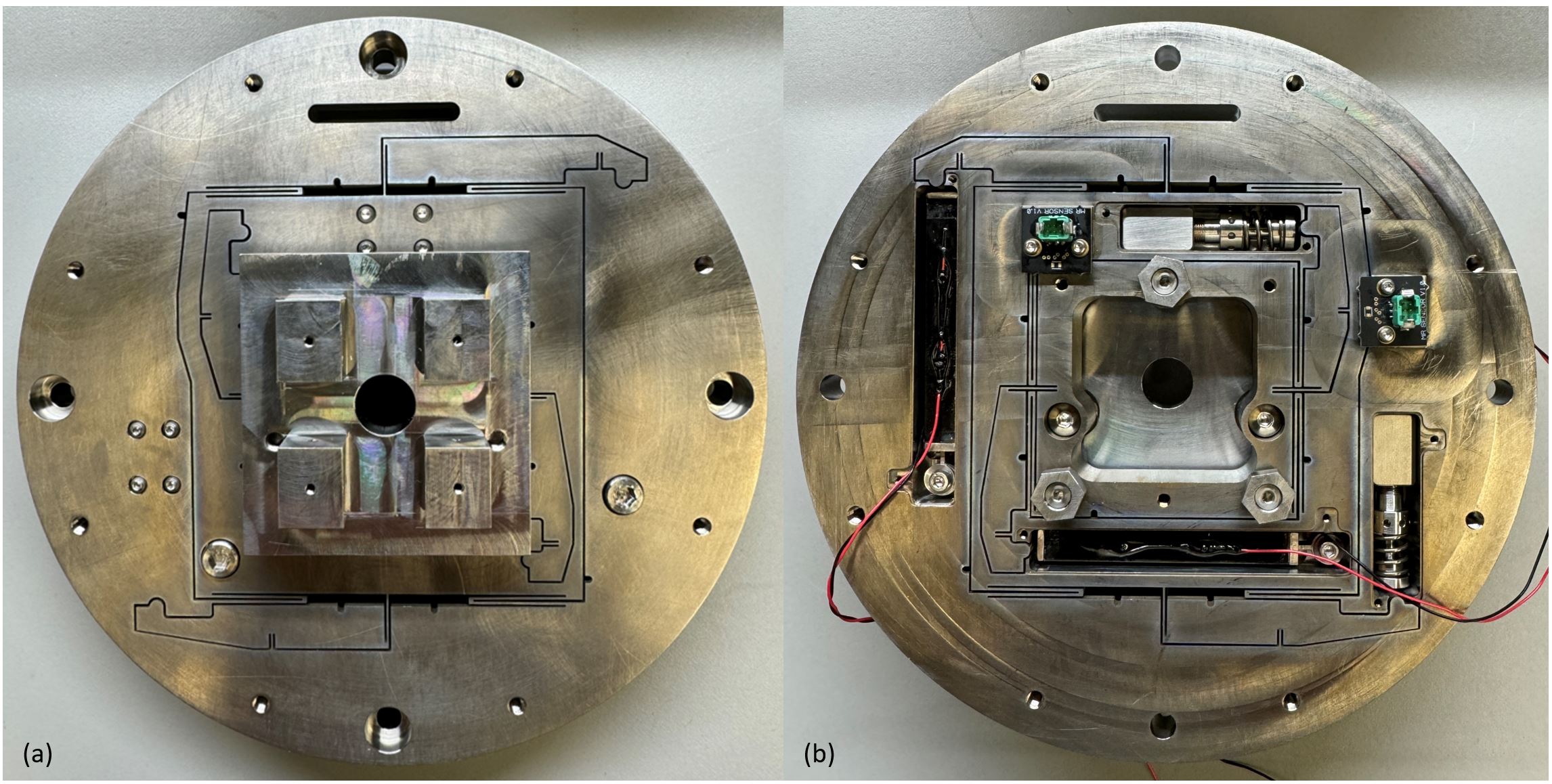}
    \caption{Photograph of the DiFCS prototype.  (a) Front side. (b) Rear side.  The two-axis nested flexure stage utilizes a piezoelectric element in each degree of freedom to provide high resolution motion for detector dithering and instrument flexure compensation. A 7:1 lever between the piezoelectric element and stage amplifies the piezo motion; necessary for adequate range of motion.  Motion encoding in each axis is provided by magnetoresistive sensors.}
    \label{fig:DiFCS_Photo}
\end{figure}

Upon receipt of the bare piezo elements from American Piezo, the devices were powered and cycled in a free state to verify the expected room temperature performance. Indeed, the stroke produced for a 150V excitation was approximately 60um, in agreement with expectation.  After initial testing of the bare elements, the Alloy 42 attachment tips were bonded to the ends of the element using 3M 2216 two-part epoxy.  Fused silica glass beads having a 100~$\mu$m diameter embedded in the epoxy establish the bond-line thickness.  Once cured, the piezos were installed into the DiFCS stage, along with the remaining mechanics, including the preload spring sub-assembly.  A photograph of the assembly is shown in Figure~\ref{fig:DiFCS_Photo}.

\begin{figure}[h]
    \centering
    \includegraphics[width=0.7\textwidth]{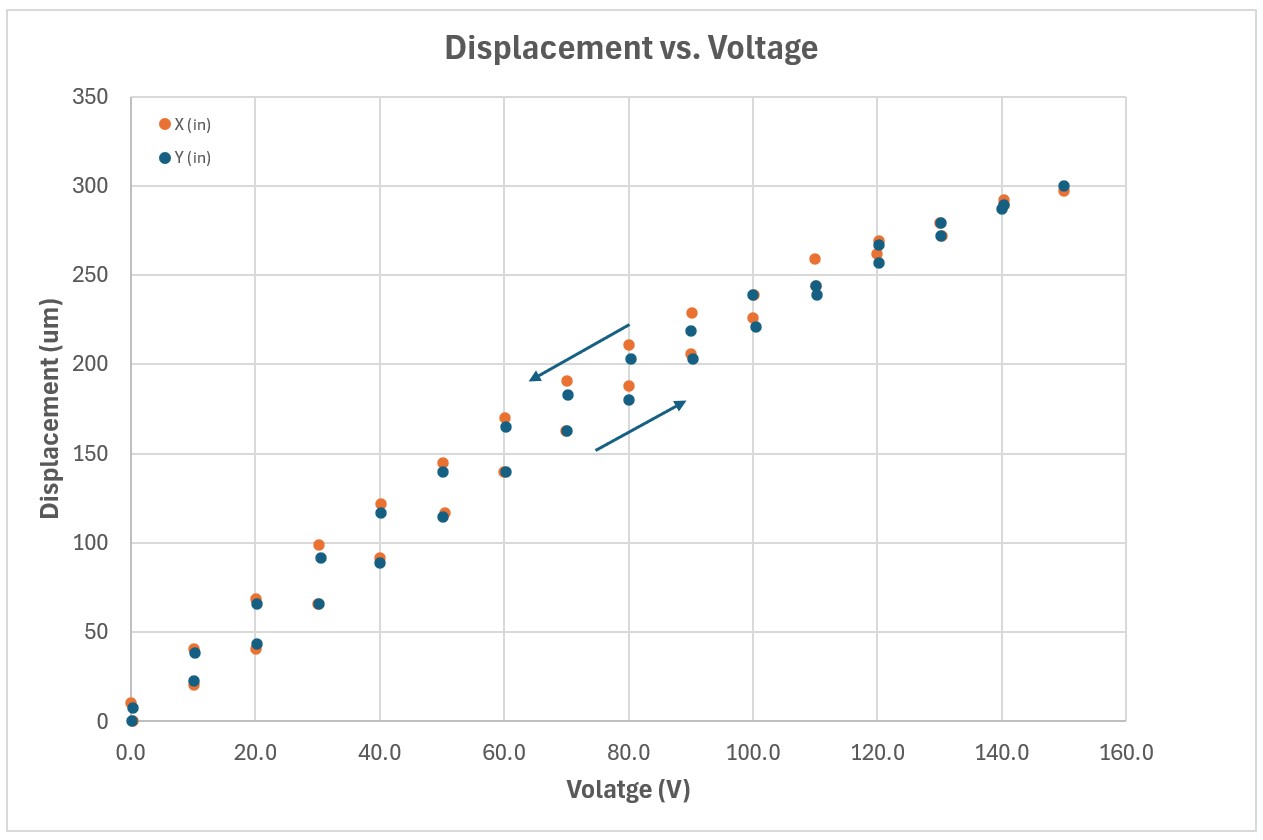}
    \caption{DiFCS stage motion as a function of applied voltage in open-loop mode.}
    \label{fig:disp_ambient}
    \vspace{1cm}
\end{figure}

Once assembled, a basic functional system test was performed at room temperature where each stage axis was exercised simultaneously. For this initial system test the magnetoresistive sensors were not operational.  Direct current voltage was applied independently to each actuator and the displacement was measured with a dial indicator. Displacement as a function of voltage for each axis is shown in Figure~\ref{fig:disp_ambient}.  It can be seen from the plot that the two axes behave in a similar fashion.  Both having the same maximum stroke and the same level of hysteresis.  

Both axes fall well short of the expected 420~$\mu$m.  Further investigation has revealed that the likely cause for the reduced stroke is the spring preload force applied to the piezo.  Despite being well below the blocking force, the preload does appear to have an impact.  Under no load, both piezos produce the expected 60~$\mu$m stroke.  However, when installed in the DiFCS stage, the compressive load on the piezo element reaches a maximum of 300~N at full travel due to the spring preload, reducing the piezo stroke is roughly 40~$\mu$m, well below expectation given that the device has a blocking force of 3500~N.  Per the manufacturer datasheet, the expected loss in stroke should be approximately 5~$\mu$m for a 300~N load. As of this writing, this issue is being actively investigated with the piezo supplier.  A possible mitigation strategy is to operate the piezos in a bi-polar mode, with an applied voltage range of -30~V to + 150~V; and at cryogenic temperature an even broader range of voltage is permissible.

\subsubsection{Cryogenic Performance}
\label{sec:cryo_meas}

As of this writing, no data has been collected from the DiFCS at cryogenic temperature.  A setup to collect this data is currently being assembled.  Unfortunately it was not completed in time for this publication, due in large part to a delay in receipt of the fiber-based interferometer procured for this test.  However, a basic test of the magnetoresistive sensors has been carried out to alleviate concerns regarding the operation of these devices well outside the stated thermal specification.  The 77~K test performed on these sensors produced very promising results.  The devices do work at cryogenic temperature, and while cryogenic operation does reduce the resistance of the devices (approximately 40\%), the decreased resistance should not impact their operation in the Wheatstone Bridge configuration. At this point we are optimistic with regard to our position-sensing strategy and will carry out full-scale testing once the cryogenic test system is complete.

\begin{figure}[h!]
    \centering
    \includegraphics[width=0.99\textwidth]{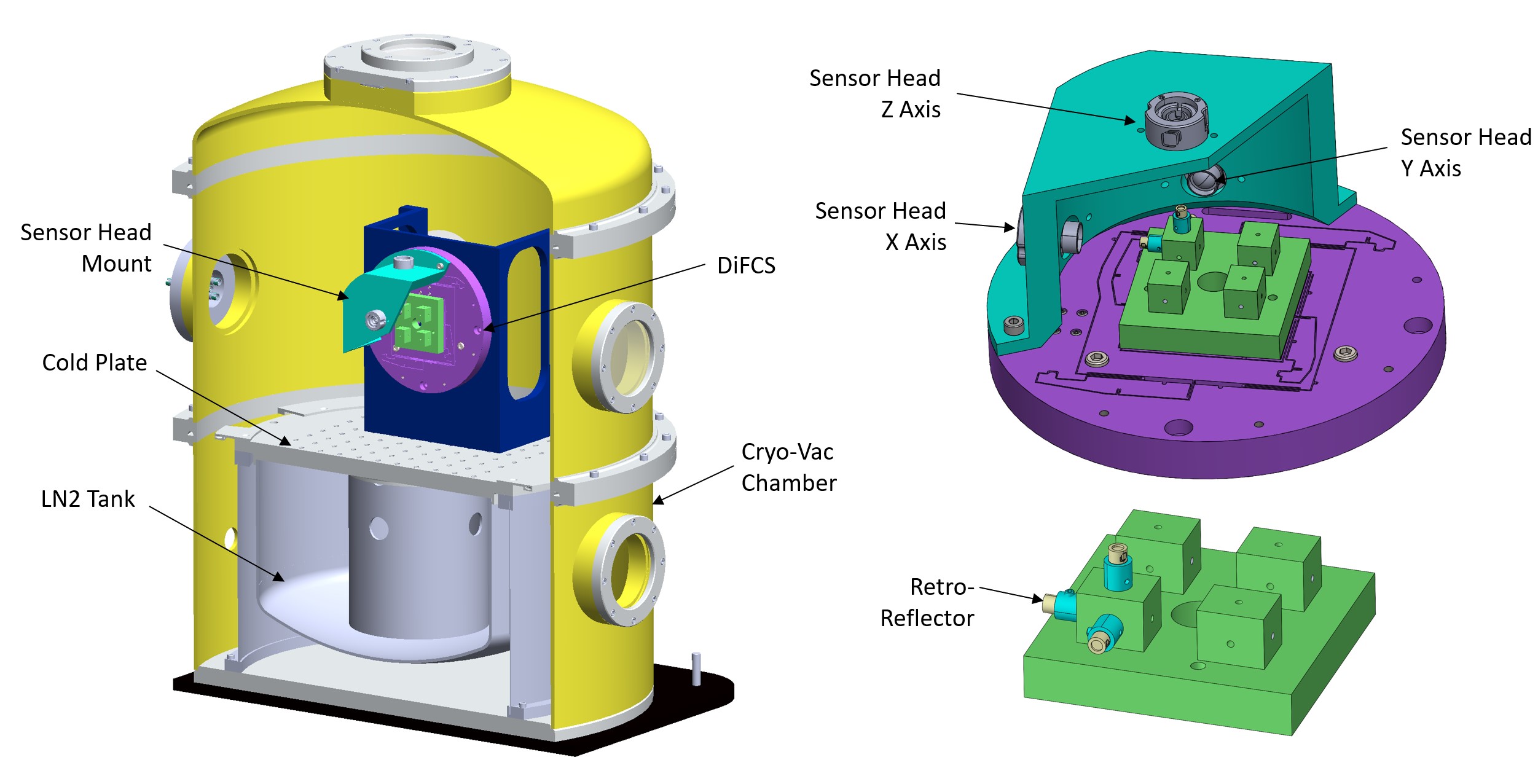}
    \caption{Metrology configuration for DiFCS characterization at cryogenic temperature.  The DiFCS is mounted to a the cold plate of the test chamber.  Sensor heads for the fiber based interferomenter are positioned to measure X, Y, and Z motion simultaneously.}
    \label{fig:cryo_setup}
\end{figure}

The setup devised to evaluate the range of motion, resolution, and accuracy of the DiFCS stage at cryogenic temperature is shown in Figure~\ref{fig:cryo_setup}.  Motion in each degree of freedom will be measured using a Fabrey Perot fiber-based interferometer system from Attocube Sytems AG.  The fiber-fed system allows the optical heads and retros to be attached locally to the DiFCS stage and operate in a vacuum and at cryogenic temperature.  Optical fibers route light to/from the three sensing heads mounted to the DiFCS out through a vacuum feedthrough in the test chamber to a three-axis real-time interferometric displacement monitor.   In this setup, X, Y, and Z motion is measured simultaneously at the corner of a mass-mockup that represents the detector and the detector base-plate.  The sensor head assembly and retros can be clocked in 90 degree increments to measure the behavior of each corner of the surrogate detector.  This metrology system provides nanometer level resolution, considerably better than required for this application.

\section{Summary}
\label{sec:summary}

A compact X-Y piezo-driven flexure detector stage for instrument flexure compensation and dithering that satisfies the volume requirements for the MIRMOS instrument has been designed and prototyped.  However, room temperature measurements show reduced stroke relative to design expectation and the root cause is currently being investigated. Given that the unloaded performance of the piezo devices meet expectation, several explanations have been explored.  Careful measurements were taken to investigate unwanted slip at the interface between the piezo tip and the stage lever.  A small adjustment was made to the design that yielded a small improvement.  An alternate power supply was used to rule out the possibility of a supply malfunction.  This led to identical results.  Lastly, measurements were made to investigate unwanted compliance between the fixed end of the piezo and the stage; a challenging measurement to make.  Results were inconclusive, but indicate no gross issue with the fixed-end connection.  

At present, the investigation into reduced stroke appears to point to the spring preload as the most likely cause; although the spring preload is 1/10th the blocking force for the device. Additional testing will be conducted in the near future to further debug this problem.  And cryogenic characterization will be carried out to assess the impact of temperature on performance; a separate, but important, matter.  As stated previously additional gains in stroke are to be had by operating the piezo in a bi-polar fashion, and by operating the device outside the ambient temperature voltage limits; something that is possible at cryogenic temperature due to the increased resistance to depoling.

\acknowledgments 
 
This material is based on substantial funding from Carnegie Science.  Funding for MIRMOS has also been graciously provided by the Heising-Simons Foundation through grant 2021-2614.

\newpage
\bibliography{report} 

\begin{thebibliography}{1}

\bibitem{2022SPIE12184E..15K}
{Konidaris}, N.~P., {Rudie}, G.~C., {Newman}, A.~B., {Lanz}, A.~E., {Williams},
  J.~E., {Hare}, T.~S., {Barkhouser}, R., {Brady}, J., {Crane}, J.~D.,
  {Kelson}, D.~D., {Killion}, G., {Kowal}, V., {Ramirez}, S., {Smee}, S.~A.,
  {Teske}, J.~K., and {Wachter}, S., ``{The Magellan infrared multi-object
  spectrograph project: 2022 update},'' in [{\em Ground-based and Airborne
  Instrumentation for Astronomy IX}{\nolinebreak\hspace{0.1em}]},  {Evans},
  C.~J., {Bryant}, J.~J., and {Motohara}, K., eds., {\em Society of
  Photo-Optical Instrumentation Engineers (SPIE) Conference Series} {\bf
  12184},  1218415 (Aug. 2022).

\bibitem{2010SPIE.7739E..3OS}
{Smee}, S.~A., ``{A precision lens mount for large temperature excursions},''
  in [{\em Modern Technologies in Space- and Ground-based Telescopes and
  Instrumentation}{\nolinebreak\hspace{0.1em}]},  {Atad-Ettedgui}, E. and
  {Lemke}, D., eds., {\em Society of Photo-Optical Instrumentation Engineers
  (SPIE) Conference Series} {\bf 7739},  77393O (July 2010).

\bibitem{2011PASP..123...87S}
{Smee}, S.~A., {Barkhouser}, R.~H., {Scharfstein}, G.~A., {Meixner}, M.,
  {Orndorff}, J.~D., and {Miller}, T., ``{Design of the WIYN High Resolution
  Infrared Camera (WHIRC)},'' {\em PASP}~{\bf 123},  87 (Jan. 2011).

\end{thebibliography}
\bibliographystyle{spiebib} 

\end{document}